\documentclass[conference]{IEEEtran}
\usepackage{cite}
\usepackage{amsmath,amssymb,amsfonts}
\usepackage{algorithmic}
\usepackage{graphicx}
\usepackage{textcomp}
\usepackage{xcolor}
\usepackage[]{mdframed}

\usepackage{amssymb}

\def\BibTeX{{\rm B\kern-.05em{\sc i\kern-.025em b}\kern-.08em
    T\kern-.1667em\lower.7ex\hbox{E}\kern-.125emX}}

\usepackage{hyperref}
\hypersetup{
    hidelinks
}
\begin{document}

\title{AI Assistants for Incident Lifecycle in a Microservice Environment — A Systematic Literature Review\\
}

\author{\IEEEauthorblockN{Dahlia Ziqi Zhou}
\IEEEauthorblockA{\textit{Department of Electrical Engineering and Computer Science} \\
\textit{York University}\\
Toronto, Canada \\
zdoris13@yorku.ca}

\IEEEauthorblockN{Marios Fokaefs}
\IEEEauthorblockA{\textit{Department of Electrical Engineering and Computer Science} \\
\textit{York University}\\
Toronto, Canada \\
mfokaefs@yorku.ca}
}

\maketitle

\begin{abstract}
Incidents in microservice environments can be costly and challenging to recover from due to their complexity and distributed nature. Recent advancements in artificial intelligence (AI) offer promising solutions for improving incident management. This paper systematically reviews primary studies on AI assistants designed to support different phases of the incident lifecycle. It highlights successful applications of AI, identifies gaps in current research, and suggests future opportunities for enhancing incident management through AI. By examining these studies, the paper aims to provide insights into the effectiveness of AI tools and their potential to address ongoing challenges in incident recovery.
\end{abstract}

\section{Introduction}
Incident management is a critical aspect of maintaining the reliability and availability of complex software systems, particularly in microservice-based architectures. As microservices continue to be widely adopted—63 \% of enterprises have transitioned to this architecture \cite{microserviceadoption}—their dynamic and distributed nature presents unique challenges. Unlike traditional monolithic systems, microservices consist of numerous independently running components that must communicate seamlessly to function as a whole. As pointed out by Liu et al. \cite{microservicechallenge}, this complexity can make debugging and repairing systems significantly more difficult. Identifying the root cause of an incident often requires tracking interactions across multiple microservices, which can be time-consuming and error-prone. Additionally, the massive volumes of data generated—logs, traces, and metrics—complicate manual analysis, making diagnosing and resolving issues quickly harder.

Managing incidents in such an environment involves navigating the four phases of the incident lifecycle: prevention, detection, containment, and post-incident analysis, as defined by the National Institute of Standards and Technology (NIST) \cite{securityguide}. In a microservice environment, the interdependent nature of services makes each phase more difficult to manage. For example, detection might require analyzing numerous data sources to identify anomalies, and containment could involve isolating failures across distributed services. As systems grow in scale and complexity, traditional methods of incident management often fall short, leading to longer downtime and increased financial losses.

For instance, a 2017 AWS outage lasting only four hours resulted in a \$150 million loss for S\&P 500 companies alone \cite{awsoutage}. Similarly, a 14-hour Facebook outage in 2020 led to an estimated \$90 million in lost revenue \cite{ccnFacebooksCatastrophic}. The Consortium for Information \& Software Quality (CISQ) estimates that the cost of poor software quality in the U.S. has reached at least \$2.41 trillion \cite{costofincident}. These examples illustrate the urgent need for more effective incident management tools and strategies to minimize downtime and its financial impact.

Recent advancements in artificial intelligence (AI) and large language models (LLMs) offer promising solutions to the challenges of managing incidents in microservice architectures. AI assistants powered by LLMs are increasingly being developed to support various phases of the incident lifecycle. These AI-driven tools can process vast amounts of unstructured data—such as logs, traces, and metrics—more efficiently than traditional methods, improving the speed and accuracy of incident detection, root cause analysis, and post-incident reporting. The ability of LLMs to analyze large datasets, generate insights, and even suggest remedial actions has opened new opportunities for automating complex incident management tasks.

Given the escalating costs of system failures and the increasing complexity of microservice-based architectures, the need for more advanced and efficient incident management tools has never been greater. AI assistants, particularly those powered by large language models, present a transformative opportunity to reshape how incidents are handled, offering the potential to not only reduce downtime but also revolutionize the entire incident lifecycle. This paper conducts a systematic literature review to provide a comprehensive understanding of the current state of AI-assisted incident management by addressing the following critical research questions:

\begin{itemize}
    \item Which phases of the incident lifecycle are these AI assistants assisting with?
    \item What are the goals of these AI assistants?
    \item What kind of methods do these AI assistants use?
    \item What type of data do these tools use to assist with incidents?
\end{itemize}

By exploring these questions, this review aims to reveal both the successes and gaps in existing research, offering crucial insights into the future of AI in incident management. The findings will highlight the current impact of AI on incident response while identifying untapped opportunities where AI can further enhance the resilience and reliability of complex software systems.

\section{Related Studies}

While no secondary studies directly address the specific focus of this paper, several related works explore topics closely aligned with microservice architecture and automated incident response.

Alshuqayran et al. \cite{microservicemapping} conducted a systematic mapping study on microservice architecture, highlighting key research areas and various diagrams used to represent these architectures. Liu et al. \cite{microservicechallenge} provided an overview of microservices and container technologies, emphasizing challenges such as performance and debugging. Li et al. \cite{microservicequality} performed a systematic literature analysis to investigate quality attributes of microservices, detailing six key attributes—including scalability, performance, and security—and corresponding mitigation strategies. Usman et al. \cite{surveyobservability} surveyed observability solutions for managing complex and distributed microservice environments, categorizing approaches from both academia and industry, and noted that while the field is evolving, current research is predominantly led by industry.

In the domain of automated incident response, Chen et al. \cite{intelligentincident} conducted an empirical study on incident management practices at Microsoft, identifying challenges such as incomplete service dependency mapping and inadequate resource health assessment. They introduced IcM BRAIN, an MLOps framework, which showed promising results in practice. Karlzen et al. \cite{incidentresponse} reviewed 45 automated solutions for security incident response, analyzing inputs and outputs through the D3FEND framework, and provided insights into state-of-the-art automated incident handling. They identified common inputs as platform monitoring and network traffic analysis, with solutions focusing on network isolation, process eviction, and platform hardening.

Regarding AI integration in DevOps, Ali et al. \cite{aidevops} reviewed literature on incorporating AI into DevOps workflows, discussing benefits, challenges, and real-world applications, and concluded that AI significantly enhances DevOps practices. Mboweni et al. \cite{mlops} conducted a systematic review of machine learning DevOps, highlighting five key themes in the literature, including model development, tool proposals, and state-of-the-art reviews.

These studies collectively deepen the understanding of microservice architecture and automated incident response, providing a foundation for exploring the role of AI assistants in these domains.

\section{Methods} 

This review adheres to the Software Engineering Guidelines for REporting Secondary Studies (SEGRESS) \cite{SEGRESS}, ensuring that the research is conducted and presented with transparency and fairness. The methodology begins by detailing the search strategy, including the formulation of the search string used to identify relevant studies. Next, the eligibility criteria for selecting papers are outlined, along with the rationale behind their inclusion or exclusion. The review also defines the key variables used to categorize and analyze each selected paper. Finally, the data collection process is explained to ensure reproducibility and clarity.

\subsection{Search Strategy} \label{sect:search-string}

Table \ref{tab:search-string} details the construction of the search string. Based on the research questions, this study created a list of search terms to narrow down the results to primary studies relevant to the paper's objectives.

\begin{table}[htbp]
\centering
\caption{Construction of the search string}
\label{tab:search-string}
\begin{tabular}{|p{0.3\linewidth} | p{0.6\linewidth} |}
\hline 
\textbf{Purpose} &  \textbf{Search Term}\\
\hline 
Context & software AND microservice* \\
\hline 
Incident lifecycle & anomal* OR bug OR event \\
\hline 
AI assistant & large language model*” OR gen* ai OR foundation model* OR LLM OR *GPT OR chatbot\\ 
\hline
Data type & trac* OR log* OR metric*\\ 
\hline
\end{tabular}
\end{table}

The search strings used in this review were carefully constructed to align with the research questions, ensuring comprehensive coverage of relevant studies:
\begin{itemize}
    \item The first term targeted studies related to microservice architectures, which were the focus of this review due to their increasing complexity and the challenges they present in incident management.
    \item The second term was directly connected to the various phases of the incident lifecycle—prevention, detection, containment, and post-incident analysis. By including terms like ``anomaly," ``bug," and ``event," the search captured papers addressing the types of issues AI assistants might help manage.
    \item The third term was designed to capture studies focused on AI assistants, particularly those leveraging advanced AI technologies like LLLMs and generative AI, which were central to the review's focus on innovative incident management solutions.
    \item The last term corresponded to the types of data commonly used in incident response (traces, logs, and metrics), addressing the research question on the data types utilized by AI assistants to support incident detection and resolution.
\end{itemize}

Each term played a critical role in identifying studies that informed the research questions on the phases of the incident lifecycle, the goals of AI assistants, the methods they employed, and the data types they utilized.

Given the novelty of the topic in software engineering, we included scientific preprints as well as published papers and journals to ensure a comprehensive overview of the incident landscape. We used data sources including IEEE, ACM, SpringerLink, Scopus, Engineering Village, and Arxiv. The first five sources are established scientific databases with extensive collections of peer-reviewed journals and papers, providing a broad search for relevant studies. Arxiv, an open-access repository for preprints awaiting peer review, was included to capture studies that have not yet been reviewed.

\subsection{Eligibility Criteria}

The review collected all results returned by the search string from each data source. The inclusion criteria listed below were necessary to narrow down the available literature to focus on studies that directly address each research question. The exclusion criteria were defined to exclude the studies that are of low quality, maintaining the overall reliability of the review findings.

\begin{enumerate}
\item Inclusion Criteria
    \begin{enumerate}
        \item Relevance: The study is relevant to the topic of discussion, focusing on how AI can assist the incident lifecycle in a microservice environment.
        \item Tool: The study proposes a tool, framework, or model that improves the incident management process.
        \item Language: The study is written in English
        \item Study Type: The study is a peer-reviewed conference paper, journal, or preprint.
    \end{enumerate}
\item Exclusion Criteria:
    \begin{enumerate}
        \item Page Length: The length of the study is less than 8 pages.
        \item Data: The study does not perform an empirical study to demonstrate the tool's effectiveness.
        \item Primary Study: The study is not a primary study.
    \end{enumerate}
\end{enumerate}

\subsection{Data Items} \label{data-item}
Table \ref{tab:variable} outlines the variables used to categorize each selected paper in response to the research questions.

The Computer Security Incident Handling Guide by the National Institute of Standards and Technology (NIST) defined four phases of the incident lifecycle: preparation (Prepare), detection and analysis (Detect), containment, eradication, and recovery (Contain), and post-incident activity (Post-incident) \cite{securityguide}. The Prepare phase focused on prevention. In the Detect phase, incidents were identified and analyzed. The Contain phase aimed to limit the impact and recover the system. The Post-incident phase involved retaining data for future learning. Papers were categorized into Prepare, Detect, Contain, and Post-incident to address RQ1.

The preliminary review indicated that many related papers focused on anomaly detection and root cause analysis. To address RQ2, the review categorized the goals of AI assistants into three main areas: anomaly detection, root cause analysis, and other categories.

For RQ3, the review classified methods into three categories: traditional methods, deep learning (DL), and LLM. Traditional methods defined machine learning techniques that rely on historical data and feature engineering, including graph-based and rule-based methods. DL methods utilize neural networks with multiple layers to automatically learn and extract features. LLMs refer to advanced deep learning models trained on extensive text corpora to understand and generate human-like language, featuring large neural networks with billions of parameters.

To address RQ4, the review focused on traces, metrics, and logs, known as the three pillars of observability \cite{sridharan2018distributed}. Studies were categorized based on the types of data used into traces, metrics, logs, and others.

\begin{table}[htbp]
\centering
\caption{List of variables used to categorize each selected paper and answer each research question}
\label{tab:variable}
\begin{tabular}{|p{0.2\linewidth} | p{0.6\linewidth} |}
\hline 
\textbf{Research Question} &  \textbf{Categories} \\
\hline 
RQ1: Incident Lifecycle & Prepare, Detect, Contain, Post-incident \\
\hline
RQ2: Goal & Anomaly detection, Root cause analysis, other \\
\hline
RQ3: Methods & Traditional Method, DL, LLM \\
\hline
RQ4: Data & Traces, Logs, Metrics, Other \\
\hline
\end{tabular}
\end{table}

\subsection{Data Collection And Analysis}

The final search string was constructed by combining the search items listed in Table \ref{tab:search-string} with AND. This search string was applied uniformly across all data sources. All papers returned by the search were collected. 

After gathering all papers returned by the databases, each paper was examined by the first author. The inclusion criteria were first applied by reviewing the abstract of each paper, including only those studies that met the criteria. The content of each paper was then reviewed, and those fitting the exclusion criteria were removed. This process was documented in detail, noting all papers returned from each data source, the reason for defining a study as irrelevant, and the criteria used for inclusion or exclusion. Please refer to \url{https://github.com/yorku-ease/SLR-incident-lifecycle-microservice} for the full dataset.

The selected papers were then examined in detail. For each research question, the author assigned the category or categories that best defined the primary study.

\section{Results}

\begin{table}[htbp]
\centering
\caption{Data Source and Query Result}
\label{tab:data-source}
\begin{tabular}{|p{0.2\linewidth} | p{0.2\linewidth} | p{0.3\linewidth} | p{0.1\linewidth} |}
\hline 
\textbf{Data Source} &  \textbf{Search Date}  & \textbf{Additional Filter} & \textbf{Search Result}\\
\hline 
IEEE & April 16 2024 & None & 21\\
\hline 
ACM & April 17 2024 & Research Article & 45 \\
\hline 
Engineering Village & April 24 2024 & None & 15 \\
\hline
Scopus & April 24 2024 & Computer Science, Engineering, Conference Paper, Book chapter & 31 \\
\hline
Springer Link & April 24 2024 & Chapter, Conference paper, research article, computer science, engineering, English & 51 \\
\hline
Arxiv & May 29 2024 & None & 146 \\
\hline
\textbf{Total} & &  & \textbf{309} \\
\hline
\end{tabular}
\end{table}

\begin{figure*}
  \centering
  \includegraphics[width=\linewidth]{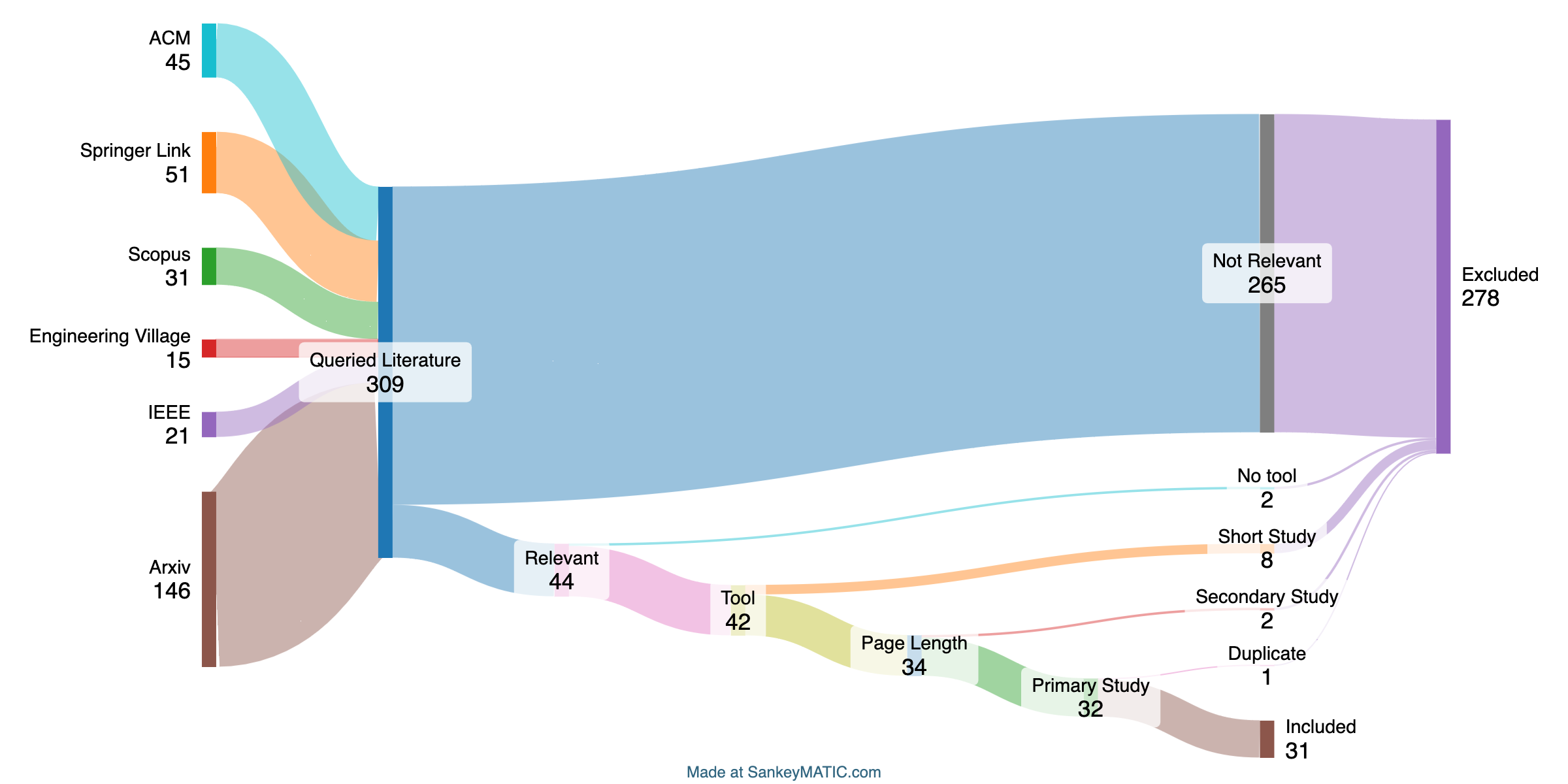}
  \caption{A Sankey Diagram of the Study Selection Process}
  \label{fig:study-selection}
\end{figure*}

\begin{table*}[htbp]
\centering
\caption{List of Selected Studies}
\label{tab:selected-studies}
\begin{tabular}{|p{0.15\linewidth} | p{0.05\linewidth} | p{0.1\linewidth} | p{0.15\linewidth} | p{0.45\linewidth} |}
\hline 
\textbf{Reference} &  \textbf{Year} & \textbf{Data Source} & \textbf{Type} & \textbf{Journal/Conference}\\
\hline 
Ma et al.\cite{knowlog} & 2024 & Scopus & Conference Paper &  IEEE/ACM International Conference on Software Engineering (ICSE)\\
\hline
Jin et al. \cite{Jin20231657} & 2023 & Scopus & Conference Paper & ACM Joint Meeting European Software Engineering Conference and Symposium on the Foundations of Software Engineering (ESEC/FSE) \\
\hline
Qi et al.\cite{Qi2023273} & 2023 & Scopus & Conference Paper & IEEE International Conference on High Performance Computing and Communications, Data Science and Systems, Smart City and Dependability in Sensor, Cloud and Big Data Systems and Application (HPCC/DSS/SmartCity/DependSys) \\
\hline
Chen et al. \cite{Chen2023359} & 2023 & Scopus & Journal & International Conference on Very Large Data Bases (VLDB) \\
\hline
Chakraborty et al. \cite{Chakraborty2023255} & 2023 & Scopus & Conference Paper & IEEE/ACM International Conference on Automated Software Engineering (ASE) \\
\hline
Lanchiano et al. \cite{Lanciano2023289} & 2023 & Scopus & Conference Paper & International Conference on Cloud Computing and Services Science (CLOSER) \\
\hline
Munir et al. \cite{logattention} & 2023 & Engineering Village & Conference Paper & International Conference on Service Oriented Computing (ICSOC) \\
\hline
Xie et al. \cite{point-wise} & 2023 & Engineering Village & Conference Paper & ACM Joint Meeting European Software Engineering Conference and Symposium on the Foundations of Software Engineering (ESEC/FSE) \\
\hline
Jiang et al. \cite{sparselog} & 2023 & Engineering Village & Conference Paper & ACM Web Conference \\
\hline
Zhang et al. \cite{tracecrl} & 2022 & Engineering Village & Conference Paper & ACM Joint Meeting European Software Engineering Conference and Symposium on the Foundations of Software Engineering (ESEC/FSE) \\
\hline
Liu et al. \cite{Liu2020} & 2020 & IEEE & Conference Paper & IEEE International Symposium on Software Reliability Engineering (ISSRE) \\
\hline
Zhang et al. \cite{Zhang2023} & 2023 & IEEE & Conference Paper & IEEE International Symposium on Software Reliability Engineering (ISSRE) \\
\hline
Chen et al. \cite{Chen2021} & 2021 & IEEE & Conference Paper & IEEE Annual Computers, Software, and Applications Conference (COMPSAC) \\
\hline
Lee et al. \cite{Lee2023} & 2023 & IEEE & Conference Paper & IEEE/ACM International Conference on Software Engineering (ICSE) \\
\hline
Wang et al. \cite{Wang2021} & 2021 & IEEE & Conference Paper & IEEE/ACM International Conference on Automated Software Engineering (ASE) \\
\hline
Wang et al. \cite{Wang2024} & 2024 & IEEE & Journal & IEEE Transactions on Reliability \\
\hline
Fukuda et al. \cite{Fukuda2022} & 2022 & IEEE & Conference Paper & IEEE/IFIP Network Operations and Management Symposium (NOMS) \\
\hline
Liu et al. \cite{liu2024largelanguagemodelsdeliver} & 2024 & Arxiv & Preprint & N/A \\
\hline
Zhang et al. \cite{zhang2024mabcmultiagentblockchaininspiredcollaboration} & 2024 & Arxiv & Preprint & N/A \\
\hline
Mulul et al. \cite{malul2024genkubesecllmbasedkubernetesmisconfiguration} & 2024 & Arxiv & Preprint & N/A \\
\hline
Zhang et al. \cite{Zhang2024Automated} & 2024 & Arxiv & Conference Paper & ACM International Conference on the Foundations of Software Engineering \\
\hline
Wang et al. \cite{wang2023rcagentcloudrootcause} & 2023 & Arxiv & Preprint & N/A \\
\hline
Huang et al. \cite{huang2024} & 2024 & Arxiv & Conference Paper & IEEE/ACM International Conference on Software Engineering: Software Engineering in Practice (ICSE-SEIP) \\
\hline
Jiang et al. \cite{jiang2024} & 2024 & Arxiv & Conference Paper & IEEE/ACM International Conference on Software Engineering (ICSE) \\
\hline
Goel et al. \cite{goel2024} & 2024 & Arxiv & Conference Paper & ACM International Conference on the Foundations of Software Engineering \\
\hline
Gupta et al. \cite{gupta2023} & 2023 & Arxiv & Conference Paper & IEEE International Conference on Cloud Computing (CLOUD) \\
\hline
Zheng et al. \cite{zheng2024mulan} & 2024 & Arxiv & Conference Paper & ACM Web Conference \\
\hline
Wang et al. \cite{wang2024fewshotcrosssystemanomalytrace} & 2024 & Arxiv & Preprint & N/A \\
\hline
Liu et al. \cite{liu2023logbasedanomalydetectionbased} & 2023 & Arxiv & Preprint & N/A \\
\hline
Zhang et al. \cite{Zhang2023robust} & 2023 & Arxiv & Journal & IEEE Transactions on Services Computing \\
\hline
Zang et al. \cite{zang2024mladunifiedmodelmultisystem} & 2024 & Arxiv & Preprint & N/A \\
\hline
\end{tabular}
\end{table*}
\subsection{Study selection}

Table \ref{tab:data-source} presents the query results from the search string defined in Section \ref{sect:search-string}. Each row corresponds to a specific data source, showing the date of the query, the additional filters applied, and the number of academic studies returned.

Figure \ref{fig:study-selection} features a Sankey diagram illustrating the study selection process, including the number of studies retrieved from each data source and those excluded based on the predefined criteria. After applying these inclusion and exclusion criteria, 31 primary studies were ultimately selected for this review. Table \ref{tab:selected-studies} provides a comprehensive list of the included primary studies.

\subsection{Study characteristics}

\begin{figure}[htbp]
  \centering
  \includegraphics[width=\linewidth]{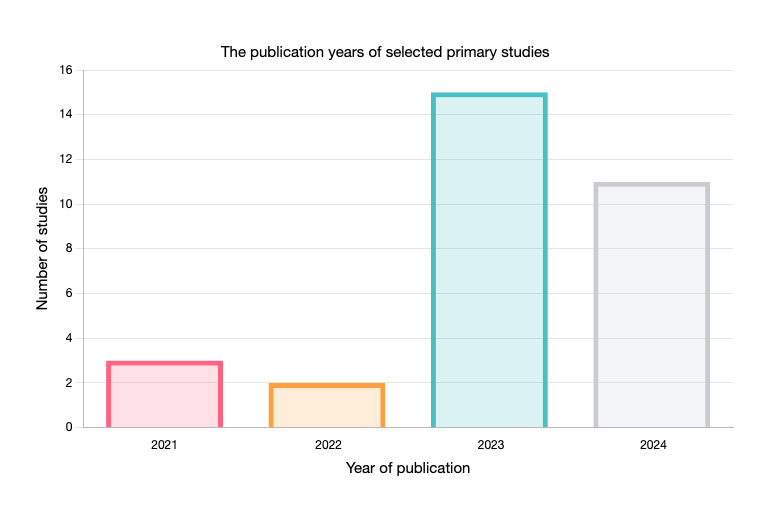}
  \caption{The Publication Years of Selected Primary Studies}
  \label{fig:publication-year}
\end{figure}

The criteria applied didn't restrict the publication year of included studies. Figure \ref{fig:publication-year} shows the distribution of selected studies' publication years. Most studies were published in 2023 and 2024. None of the studies was published before 2021. This distribution indicates that using AI assistant for microservice environment incidents has been a new research topic in recent years. 

Table \ref{tab:result} contains the list of included primary studies and their characteristics. Each row represents a primary study. A checkmark under a specific column indicates the study has the respective characteristic. Section \ref{sect:result} analyzes this table in detail to answer the research questions.

\subsection{Results of analyses} \label{sect:result}

\begin{table*}[htbp]
\centering
\caption{Overview of selected primary studies}
\label{tab:result}
\begin{tabular}{|p{0.045\linewidth} | p{0.045\linewidth} | p{0.045\linewidth} | p{0.045\linewidth} | p{0.045\linewidth} | p{0.05\linewidth} | p{0.045\linewidth} | p{0.045\linewidth} | p{0.05\linewidth} | p{0.04\linewidth} | p{0.04\linewidth} | p{0.035\linewidth} | p{0.035\linewidth} | p{0.035\linewidth} | p{0.035\linewidth} |}
\hline 
\multicolumn{1}{|p{0.045\linewidth}|}{\textbf{Primary Study}} & \multicolumn{4}{|p{0.18\linewidth}|}{\textbf{Incident Lifecycle}} & \multicolumn{3}{|p{0.14\linewidth}|}{\textbf{Goals}} & \multicolumn{3}{|p{0.13\linewidth}|}{\textbf{Methods}} & \multicolumn{4}{|p{0.14\linewidth}|}{\textbf{Data}}\\
\hline 
& Prepare & Detect & Contain & Post-incident & Anomaly Detection & Root Cause Analysis& Others & Tradition Method & DL & LLM & Traces & Logs & Metrics & Others \\

\hline 
\cite{knowlog} & & \checkmark  & & & & & \checkmark & && \checkmark & & \checkmark  & &  \\
\hline 
\cite{Jin20231657} & & \checkmark & & & & & \checkmark & \checkmark & \checkmark & \checkmark & & & & \checkmark  \\
\hline 
\cite{Qi2023273} & & \checkmark & & & \checkmark & & & & & \checkmark & & \checkmark & &  \\
\hline 
\cite{Chen2023359} & & \checkmark & & & \checkmark & & & & \checkmark & & & & & \checkmark \\
\hline 
\cite{Chakraborty2023255} & & & \checkmark & & & \checkmark & &\checkmark & & & & & & \checkmark \\
\hline 
\cite{Lanciano2023289} &\checkmark & & & & & & \checkmark & & & \checkmark & & & & \checkmark \\
\hline 
\cite{logattention} &\checkmark & & & & \checkmark & & & & \checkmark & & & \checkmark & &  \\
\hline 
\cite{point-wise} & & \checkmark & & & \checkmark & & & \checkmark & \checkmark & & \checkmark & & &  \\
\hline 
\cite{sparselog} & & \checkmark & & & \checkmark & & & \checkmark & & & & \checkmark & &  \\
\hline
\cite{tracecrl} & & \checkmark & & & \checkmark & & & & \checkmark & & \checkmark & & &  \\
\hline
\cite{liu2023logbasedanomalydetectionbased} & & \checkmark & & & \checkmark & & & & \checkmark & & \checkmark & & &  \\
\hline
\cite{Zhang2023robust} & & \checkmark & & & \checkmark & & & \checkmark & \checkmark & & \checkmark & & &  \\
\hline
\cite{Chen2021} & & \checkmark & & & & \checkmark & & & \checkmark & & \checkmark & & &  \\
\hline
\cite{Lee2023} & & \checkmark & \checkmark & & \checkmark & \checkmark & & \checkmark & \checkmark & & \checkmark & \checkmark & \checkmark &  \\
\hline
\cite{Wang2021} & & & \checkmark & & & \checkmark & & \checkmark & & & & \checkmark & \checkmark & \checkmark \\
\hline
\cite{Wang2024} & & & \checkmark & & & \checkmark & & \checkmark & \checkmark & & \checkmark & & \checkmark & \\
\hline
\cite{Fukuda2022} & & \checkmark & & & & & \checkmark & & \checkmark & & & \checkmark & \checkmark &  \\
\hline
\cite{liu2024largelanguagemodelsdeliver} & & \checkmark & & & \checkmark & & & \checkmark & & \checkmark & & & \checkmark & \checkmark \\
\hline
\cite{zhang2024mabcmultiagentblockchaininspiredcollaboration} & & & \checkmark & & & \checkmark & & & & \checkmark & & & & \checkmark \\
\hline
\cite{malul2024genkubesecllmbasedkubernetesmisconfiguration} & \checkmark & & & & & & \checkmark & & & \checkmark & & & & \checkmark \\
\hline
\cite{Zhang2024Automated} & & & \checkmark & & & \checkmark & & & & \checkmark & & & & \checkmark \\
\hline
\cite{wang2023rcagentcloudrootcause} & & & \checkmark & & & \checkmark & & & & \checkmark & & \checkmark & & \checkmark \\
\hline
\cite{huang2024} & & & & \checkmark & & & \checkmark & & & \checkmark & & & & \checkmark \\
\hline
\cite{jiang2024} & &  &  \checkmark & & & & \checkmark & & & \checkmark & & & & \checkmark \\
\hline
\cite{goel2024} & & & \checkmark & & & \checkmark  & & & & \checkmark & & & \checkmark & \checkmark \\
\hline
\cite{gupta2023} & & \checkmark & & & & & \checkmark & & & \checkmark & & \checkmark & & \\
\hline
\cite{zheng2024mulan} & & & \checkmark & & & \checkmark & & & \checkmark & & & \checkmark & \checkmark & \\
\hline
\cite{wang2024fewshotcrosssystemanomalytrace} & & \checkmark & & & \checkmark & & & & \checkmark & & \checkmark & \checkmark & & \\
\hline
\cite{liu2023logbasedanomalydetectionbased} & & \checkmark & & & \checkmark &  & & \checkmark & & \checkmark & & \checkmark & & \\
\hline
\cite{Zhang2023} & & & \checkmark & & & \checkmark & & & \checkmark & & \checkmark & \checkmark & \checkmark & \\
\hline
\cite{zang2024mladunifiedmodelmultisystem} & & \checkmark & & & \checkmark & & & \checkmark & & & & \checkmark &  & \\
\hline

\end{tabular}
\end{table*}

\subsubsection{RQ1} \label{sect:rq1}

Which phases of the incident lifecycle are these AI assistants assisting with?

Based on the data presented in Table \ref{tab:result}, 3 papers described AI assistants that supported the Prepare phase. Lanciano et al. \cite{Lanciano2023289} introduced an assistant designed to identify configuration issues in declarative code for infrastructure-as-code tools. Similarly, Malul et al. \cite{malul2024genkubesecllmbasedkubernetesmisconfiguration} developed a method to prevent misconfigurations in Kubernetes files.

For the Detect phase, 17 papers focused on AI assistants that identified production issues with minimal or no human intervention. Tools such as ImDiffusion \cite{Chen2023359} and MicroCU \cite{sparselog} were specifically designed to handle challenging data types, like multivariate time series and sparse API logs. Other assistants innovated in processing the vast amounts of data generated by large microservice systems. For example, Zhang et al. \cite{Zhang2023robust} utilized traces to detect production issues, Liu et al. \cite{liu2024largelanguagemodelsdeliver} proposed an assistant for cloud outage warnings based on metrics, and Zang et al. \cite{zang2024mladunifiedmodelmultisystem} developed a model for log anomaly detection across multiple systems.

In the Contain phase, 11 AI assistants focused primarily on root cause analysis. Groot \cite{Wang2021} and Mulan \cite{zheng2024mulan} used graph-based approaches to identify the point of failure. Other researchers, such as Zhang et al. \cite{zhang2024mabcmultiagentblockchaininspiredcollaboration} and Wang et al. \cite{wang2023rcagentcloudrootcause}, leveraged past incidents and experiences to determine the root cause of new incidents. Interestingly, Jiang et al. \cite{jiang2024} developed a query assistant tool for On-call Engineers that filtered telemetry data, facilitating faster incident resolution. 

The Post-incident phase saw only one AI assistant, FaultProfIT \cite{huang2024}, which automatically categorized large numbers of incident tickets to reduce manual effort and enhance postmortem analysis.

\begin{mdframed}
    \textbf{Answer to RQ1}: The majority of AI assistants focused on the Detect and Contain phases, representing 54.8\% and 35.4\% of the selected papers, respectively. Only 9.7\% of assistants targeted the Prepare phase, while 3.2\% proposed tools that aided in the Post-incident phase.
\end{mdframed}

\subsubsection{RQ2}

What are the goals of these AI assistants?

According to Table \ref{tab:result}, 13 AI assistants in the selected primary studies were tasked with anomaly detection. For example, Qi et al. \cite{Qi2023273} explored the use of ChatGPT to process logs for identifying cloud issues. Xie et al. \cite{point-wise} investigated group-based trace anomaly detection to improve detection accuracy. Liu et al. \cite{liu2023logbasedanomalydetectionbased} developed an unsupervised method to identify trace anomalies.

Many other AI assistants focused on root cause analysis. In the selected papers, 11 assistants concentrated on this task. ESRO \cite{Chakraborty2023255} used past outage knowledge to construct a knowledge graph that ranked possible root causes. mABC \cite{zhang2024mabcmultiagentblockchaininspiredcollaboration}, inspired by blockchain, identified root causes in a microservice architecture. RCAgent \cite{wang2023rcagentcloudrootcause} developed a tool-assisted LLM assistant that utilized both code and log analysis.

Notably, Eadro \cite{Lee2023} was an end-to-end AI assistant that aided anomaly detection and root cause analysis. It monitored logs, metrics, and traces to detect potential anomalies and created a dependency graph based on the traces to identify the root cause.

The remaining AI assistants had various goals: incident summarization, improving data understanding for downstream tasks, and detecting deployment misconfigurations. Oasis \cite{Jin20231657} used relevant past incidents to generate an incident summary and impact report. Fukuda et al. \cite{Fukuda2022} produced fault reports based on logs and metrics. Both Knowlog \cite{knowlog} and the assistant developed by Gupta et al. \cite{gupta2023} aimed to improve log analysis for downstream tasks, such as log parsing and anomaly detection.

\begin{mdframed} \textbf{Answer to RQ2:} Most AI assistants aimed at anomaly detection, representing about 41.9\%, while 35.5\% focused on root cause analysis. The remaining 23.5\% of AI assistants addressed other areas to improve the incident response process. 
\end{mdframed}

\subsubsection{RQ3}

What kind of methods do these AI assistants use?

Although the search string focused on AI assistants utilizing LLMs, many of the selected primary studies presented AI assistants that employed traditional machine learning, deep learning, or a combination of multiple methods.

Among the selected primary studies, 12 AI assistants utilized LLMs. For example, GenKubeSec \cite{malul2024genkubesecllmbasedkubernetesmisconfiguration} used an LLM fine-tuned on a large and diverse set of Kubernetes configuration files, followed by files labelled with misconfigurations. Equipped with configuration prediction, the assistant identified misconfigurations and recommended remediation steps. RCAgent \cite{wang2023rcagentcloudrootcause} employed tool-augmented LLMs that utilized data-querying functions such as SQL interfaces and log query APIs, as well as LLM-based analytical agents that extended domain knowledge. These tools enabled the model to identify the root cause of incidents in real-time by accessing additional context through zero-shot prompting. LLMAD \cite{liu2024largelanguagemodelsdeliver} used in-context learning by retrieving normal and abnormal time-series patterns, along with chain-of-thought prompting to enhance performance and provide detailed explanations.

DL methods were used by 10 AI assistants. For example, TraceSieve \cite{Zhang2023robust} applied an unsupervised trace anomaly detection method based on a Variational Graph Autoencoder and analyzed a data structure called a trace feature matrix to identify trace anomalies. DiagFusion \cite{Zhang2023} encoded traces, logs, and metrics and utilized a graph neural network to diagnose failure types in real-time.

Traditional machine learning methods were used by 3 AI assistants. ESRO \cite{Chakraborty2023255}, for example, constructed a causal graph based on past alerts and a knowledge graph from past outage reports. By grouping similar outages into alerts and retrieving them through cluster-based inference, the assistant presented root-cause solutions based on similar past outages.

Six of the AI assistants employed a combination of methods. For instance, Wang et al. \cite{Wang2024} divided the system into three layers and used a transformer-based anomaly detector to generate an anomaly score for each container. The assistant also used a breadth-first search algorithm to discover the root cause area. Oasis \cite{Jin20231657} used graph methods to link incidents, neural network models to find related incidents, and a fine-tuned GPT-3 model to generate incident summaries.

\begin{mdframed} 
\textbf{Answer to RQ3:} Out of the selected AI assistants, 38.7\% employed LLMs, 32.2\% used deep learning methods, and 9.68\% utilized traditional machine learning techniques. The remaining 19.4\% of AI assistants employed a combination of multiple methods. 
\end{mdframed}

\subsubsection{RQ4}

What type of data do these AI assistants use to assist with incidents?

This review focused on the three pillars of observability: traces, logs, and metrics \cite{sridharan2018distributed}. Traces capture the path of a request as it travels through various services, helping to visualize and diagnose complex interactions in distributed systems. Logs are records of events or activities that occur within the system, providing detailed insight into what happened at specific points in time. Metrics are numerical measurements that track the performance and health of different components, such as response times, error rates, and resource usage.

According to the overview in Table \ref{tab:result}, 9 AI assistants used traces to assist with incidents, 15 AI assistants utilized logs to achieve their goals, and 8 of them worked with metrics. For example, TraceCRL \cite{tracecrl} constructed an operation invocation graph for each trace and used this representation to identify anomalous traces. BertOps \cite{gupta2023} proposed a model that was fine-tuned for AIOps using logs from 17 different data sources to improve downstream log analysis tasks. Liu et al. \cite{liu2024largelanguagemodelsdeliver} presented an assistant that processed metrics to identify metric anomalies.

In addition to these three data types, 13 assistants used other data, such as incident summaries and code, as in the case of RCAgent \cite{wang2023rcagentcloudrootcause}, telemetry queries \cite{jiang2024}, developer activities \cite{Wang2021}, and deployment configuration files \cite{Lanciano2023289}.

\begin{mdframed} \textbf{Answer to RQ4:} 
Among the selected AI assistants, 29\% used traces, 48.4\% utilized logs, and 25.8\% worked with metrics. Beyond these three common data types, 41.9\% of the assistants also made use of other data, such as incident summaries, code, and deployment configuration files. 
\end{mdframed}

\section{Discussion}

Incident management in large-scale microservice systems has attracted considerable research attention. This review synthesizes key studies that explore the potential of AI-powered assistants to enhance the various phases of the incident lifecycle.

\subsection{Research Gap}

As discussed in Section \ref{sect:result}, most studies have concentrated on the Detect phase, with a significant focus on anomaly detection using traces, metrics, logs, and other observability data. AI assistants are particularly effective at processing the large volumes of data generated by microservice systems, much of which is difficult for engineers to analyze manually. Additionally, many AI assistants featured in the primary studies also support the Contain phase, especially in conducting root cause analysis.

However, there is a notable research gap in the Prepare and Post-Incident phases. In the Prepare phase, only a few studies—such as Lanciano et al. \cite{Lanciano2023289} and Malul et al. \cite{malul2024genkubesecllmbasedkubernetesmisconfiguration}—have focused on challenges related to declarative infrastructure configuration. Munir et al. \cite{logattention} explored methods for assessing software releases to prevent production failures. According to the Incident Handler's Handbook by the SANS Institute, the Prepare phase is critical to the incident lifecycle, playing a key role in preventing failures and shaping a team's incident response readiness \cite{sansIncidentHandleraposs}. Despite its importance, significant gaps remain in this phase. Similarly, the Post-Incident phase has received limited attention. Huang et al. \cite{huang2024} is the only study proposing a tool for automating incident categorization to reduce the effort required for post-mortem analysis.

The reasons behind these research gaps may be due to the nature of the work in these phases. The Prepare phase involves tasks such as setting up tooling, creating handbooks, and defining processes—activities that are often less structured and require significant human input. Incident prevention often means addressing complex, poorly defined problems, such as improving release processes and observability practices. Similarly, post-incident learning is often neglected, even though it is considered one of the most important phases, as noted in the Computer Security Incident Handling Guide by NIST \cite{securityguide}. Atlassian, a leading issue-tracking company, suggests this neglect may stem from the perception that once an incident is resolved, there is little need for further reflection \cite{atlassianImportanceIncident}. The focus tends to be on resolving incidents rather than learning from them, as post-mortems are sometimes viewed as formalities. Thorough post-incident reviews are crucial to preventing similar issues from recurring, yet research remains limited on how AI can improve this process. Addressing the challenges in these two phases not only broadens the scope of research but also highlights the potential of AI assistants to optimize the entire incident lifecycle in microservice environments.

\subsection{Data Sources}

The three pillars of observability—traces, logs, and metrics—have long been used to understand system performance and identify degradation\cite{sridharan2018distributed}. However, this review reveals that other, less traditional data sources also hold valuable information for incident management. For instance, past incident reports have been leveraged by Zhang et al.\cite{Zhang2024Automated}, Jiang et al.\cite{jiang2024}, and others to identify patterns and perform predictive analysis. These reports often include details such as the initial alert trigger, associated logs and traces, and the steps taken during incident mitigation. This wealth of information is crucial not only during ongoing incidents but also for post-incident analysis. In a microservice environment, where the sheer volume of incident reports can be overwhelming, an AI assistant capable of processing this information becomes invaluable.

AI assistants like ESRO\cite{Chakraborty2023255} and mABC\cite{zhang2024mabcmultiagentblockchaininspiredcollaboration} also make use of alert information. Alerts, set by engineers to trigger when specific conditions indicative of failures are met, contain descriptions that can be key to identifying root causes. By comparing the sequence of alert events to past incidents, AI can effectively utilize prior knowledge to enhance incident response.

Another valuable yet underutilized data type is the service dependency graph, which maps how components within a microservice architecture interact. Goel et al.\cite{goel2024} and Groot \cite{Wang2021} both employed dependency graphs to assist with anomaly detection and root cause analysis. Groot\cite{Wang2021} further expanded this by incorporating developer activities—such as code deployments or configuration updates—into an event graph. Tracking these developer actions is essential, as they can have unexpected impacts on the system.

Code repositories, though rich in information, are often overlooked in incident management due to their large volume. However, tools like RCAgent\cite{wang2023rcagentcloudrootcause} provide code analysis capabilities that enable AI models to gain additional context, thereby improving the accuracy of root cause analysis.

\subsection{User Study}

All the primary studies conducted comprehensive quantitative evaluations to assess the effectiveness of their AI assistants, yet only 5 studies included a user study. Jin et al. \cite{Jin20231657} presented the output of OASIS, generated by different models, to 54 on-call engineers who were asked to rank their preferences. Their favorable responses towards OASIS output indicate that AI assistants can be effective in practice. Groot \cite{Wang2021} was implemented in a production environment to visualize dependency graphs, enabling incident responders to quickly understand incidents and identify root causes. Feedback from Groot users and maintainers not only demonstrated the success of its adoption but also provided valuable insights.

Liu et al. \cite{liu2024largelanguagemodelsdeliver} assessed the quality of anomaly explanations by inviting 5 experienced DevOps engineers to evaluate the outputs. Similarly, Zhang et al. \cite{zhang2024mabcmultiagentblockchaininspiredcollaboration} invited 10 AIOps experts to rate the root cause analysis and resolution capabilities of their assistant. The feedback gathered from these experts offered critical insights into the assistant’s usefulness and the best methodologies to employ. Zhang et al. \cite{Zhang2024Automated} also sought feedback from 18 incident owners to verify the accuracy and readability of the root cause analyses generated by their AI assistant. Although there was room for improvement, the feedback was largely positive.

Integrating AI assistants into production environments and conducting user studies with subject matter experts is essential for gathering valuable feedback. The insights from these experts can significantly enhance the effectiveness of AI assistants. Moreover, human feedback serves as a crucial confirmation of the tool’s practical utility, complementing quantitative evaluations and ensuring that the AI solutions meet real-world needs.

\subsection{Future opportunities}

Future research in the field of AI-assisted incident management for microservice environments presents several promising opportunities. Given the current focus on the Detect and Contain phases, there is a significant need for more research exploring AI’s potential in the Prepare and Post-Incident phases. Developing AI tools that address the unique challenges of microservice environments—such as infrastructure configuration, resource allocation, and post-mortem analysis—could lead to more proactive and comprehensive incident management strategies. Additionally, while traditional data types like traces, logs, and metrics have been widely studied, future research should explore the integration of non-traditional data sources, such as past incident reports, service dependency graphs, developer activities, and code repositories, to enrich the AI’s decision-making capabilities.

Furthermore, the importance of user studies cannot be overstated. Future research should incorporate more robust user studies alongside quantitative evaluations to ensure that AI assistants are not only effective in theory but also practical and user-friendly in real-world settings. Engaging with subject matter experts during the development and implementation phases will provide critical feedback that can refine AI tools, making them more aligned with the needs of on-call engineers, DevOps teams, and incident responders. By addressing these gaps, future research can contribute to the creation of more intelligent, responsive, and holistic AI-powered incident management systems that fully harness the potential of AI in microservice environments.

\subsection{Limitations}

This paper adheres to the Software Engineering Guidelines for Reporting Secondary Studies (SEGRESS) \cite{SEGRESS}. However, despite these methods, certain limitations of this systematic literature review should still be acknowledged.

First, selection bias was a potential concern in this review, as the inclusion criteria favoured studies that specifically reported positive outcomes or successful applications of AI assistants. To mitigate this bias, the review utilized a comprehensive search strategy across multiple databases and included a wide range of studies that met the defined criteria. However, studies with negative or inconclusive results were not considered, which might limit the scope of the findings. The review aimed to reduce bias by applying consistent inclusion and exclusion criteria and by reviewing a broad spectrum of literature to ensure a balanced overview of the current state of research.

Although the review includes preprint studies, capturing research before peer review, there remains a time lag between when research is conducted and when it becomes available. As a result, the review may not fully reflect the latest technical advances in AI, especially in such a rapidly evolving field.

Lastly, the review is limited to studies published in English, potentially excluding valuable research from non-English speaking regions. This language bias may limit the global applicability and generalizability of the findings.

In conclusion, the generalizability of the results is constrained by several factors. The review is limited to English-speaking research, which may omit relevant studies published in other languages. Additionally, it focuses specifically on microservice architectures and the application of large language models (LLMs) for incident management, potentially restricting the applicability of the findings to other software architectures and AI techniques. Furthermore, the emphasis on studies that report successful applications of AI assistants may narrow the scope, overlooking instances where these tools did not meet expectations. These limitations should be considered when interpreting the results and applying them to broader contexts.

\bibliographystyle{plain}
\bibliography{refs}

\end{document}